\documentclass[twocolumn,aps,showpacs,floatfix,prc]{revtex4}
\usepackage[dvips]{epsfig}
\begin{document}

\title{ Cluster radioactivity in very heavy nuclei: a new perspective }

\author{ T.R. Routray$^1$, Jagajjaya Nayak$^1$ and D.N. Basu$^2$ }

\affiliation{$^1$P.G. Department of Physics, Sambalpur University, Jyoti Vihar, Burla, Orissa-768019, India }
\affiliation{$^2$Variable  Energy  Cyclotron  Centre, 1/AF Bidhan Nagar, Kolkata 700 064, India }
\email[E-mail: ]{trr1@rediffmail.com}
\email[E-mail: ]{jagat.su_ph@yahoo.in}
\email[E-mail: ]{dnb@veccal.ernet.in}

\date{\today }

\begin{abstract}

    Exotic cluster decay of very heavy nuclei is studied using the microscopic nuclear potentials obtained by folding density dependent M3Y effective interaction with the densities of the cluster and the daughter nuclei. The microscopic nuclear potential, Coulomb interaction and the centrifugal barrier arising out of spin-parity conservation are used to obtain the potential between the cluster and the daughter nuclei. Half life values are calculated in the WKB framework and the preformation factors are extracted. The latter values are seen to have only a very weak dependence on the mass of the emitted cluster. 

\end{abstract}

\maketitle

\noindent
\label{section1}

    The exotic phenomena cluster radioactivity in very heavy nuclei was suggested by Sandulescu et al. \cite{Sa80} in 1980. Since the first experimental observation of cluster radioactivity \cite{Ro84}, a lot of efforts have gone into the understanding of the physics of cluster radioactivity. Predictions for $\alpha$ and various exotic decays have been made by the analytical superasymmetric fission model (ASAFM) \cite{PG86,Po91} with reasonable success. This was followed by the preformed cluster model (PCM) calculations which is distinguished by the inclusion of the cluster preformation probability and was applied to $\alpha$ decay \cite{bmp93} with similar success. But  both the theoretical approaches described above use phenomenological potentials for the nucleus-nucleus interactions. The ASAFM uses a parabolic potential approximation for the nuclear interaction within a superasymmetric fission model description which yields analytical expressions for the decay lifetimes, while the PCM uses a cos-hyperbolic form for nuclear interaction potential arising from the folding integral with sufficiently well known charge densities and the resulting nuclear potential could be more or less accurately parameterized to such a form. 

    In the present work the phenomena of nuclear cluster radioactivity is studied theoretically using microscopic potentials within WKB framework of quantum tunneling. The microscopic nuclear interaction potential is calculated by folding the density distributions of the emitted and daughter nuclei with density dependent M3Y effective interaction (DDM3Y). The DDM3Y effective nucleon-nucleon (NN) interaction used here was peviously used successfully for elastic and inelstic scattering of protons \cite{Gu05,Gu06}, proton radioactivity \cite{Ba05} and $\alpha$ radioactivity \cite{prc06,prc07} whose density dependence was obtained from nuclear matter calculations \cite{bcs08}. The cluster preformation factors are extracted from the calculated and the measured half lives of cluster radoactivity and its systematics are studied.

\noindent
\label{section2}

    The decay constant $\lambda$ for cluster radioactivity is a product of cluster preformation probability $P_0$ in the ground state, the tunneling probability through barrier $P$ and the assault frequency $\nu$. Cluster emission half life $T_{1/2}=\ln2/\lambda$ calculated within WKB framework for barrier penetration probability $P$ is:

\begin{equation}
T_{1/2} = [(h \ln2) / (2 E_v)] [1 + \exp(K)]
\label{seqn1}
\end{equation}
\noindent
where $E_v$, the zero point vibration energy, is assumed to be proportional to $Q$-value  of the spontaneous cluster emission. The WKB action integral is given by

\begin{equation}
 K=\frac{2}{\hbar}\int_{R_a}^{R_b} {\sqrt{2\mu (E(R) - E_v - Q)}} dR
\label{seqn2}
\end{equation}
\noindent
where the total interaction energy between the emitted cluster and the daughter nucleus $E(R)$=$V_N(R) + V_C(R) + \hbar^2 l(l+1) / (2\mu R^2)$ is equal to the sum of the nuclear interaction energy, Coulomb interaction energy and the centrifugal barrier arising from spin-parity conservation. The reduced mass $\mu$ = $mA_e A_d/A$ where $A_e, A_d, A$ are the mass numbers of the emitted cluster, the daughter and the parent nuclei respectively and $m$ is the nucleonic mass measured in the units of MeV/c$^2$. $R_a$ and $R_b$ are the second and third turning points of the WKB action integral determined from the condition: $E(R_a)  = Q + E_v =  E(R_b)$ which provides three classical turning points. The energetics allow spontaneous emission of clusters only if the released energy $Q = M - ( M_e + M_d)$ is a positive quantity, where $M$, $M_e$ and $M_d$ are the atomic masses (in MeV) of the parent nucleus, the emitted cluster and the residual daughter nucleus, respectively. The microscopic nuclear potentials $V_N(R)$ are obtained by double folding the densities of the emitted cluster and the residual daughter nucleus with the finite range realistic density dependent M3Y effective interacion as

\begin{equation}
 V_N(R) = \int \int \rho_1(\vec{r_1}) \rho_2(\vec{r_2}) v(s) d^3r_1 d^3r_2 
\label{seqn3}
\end{equation}
\noindent
where $s$=$|\vec{r_2} - \vec{r_1} + \vec{R}|$, $\rho_1$ and $\rho_2$ are the density distributions for the two composite nuclear fragments having the spherically symmetric form given by $\rho_{1,2}(r)$ = $\rho_0 / [ 1 + \exp( (r-c) / a ) ]$ where $c$ = $r_\rho ( 1 - \pi^2 a^2 / 3 r_\rho^2 )$, $r_\rho$ = $1.13 A_{d,e}^{1/3}$, $a$ = 0.54 fm and the value of $\rho_0$ is fixed by equating the volume integral of the density distribution function to the mass number of the fragment. The experimental charge density distributions in case of the heavier nuclei can be well described by the two parameter Fermi function \cite{Fo69} and since the charge {\it i.e.} the proton (p) and the neutron (n) density distributions should have similar forms due to the same strengths of the n-n and p-p nuclear forces, the matter density distribution can be well described by the spherically symmetric Fermi function. The density distribution function in case of $\alpha$ particle has the Gaussian form $\rho_2(r) = 0.4229 \exp( - 0.7024 r^2)$ whose volume integral equals $A_\alpha (=4)$, the mass number of the $\alpha$ particle. 

    The density dependent M3Y \cite{Be77} effective interaction (DDM3Y) \cite{Sa79} appearing in Eq.(3) is given by $v(s)$ = $t^{M3Y}(s,E)g(\rho_1,\rho_2)$ where

\begin{eqnarray}
  t^{M3Y} = 7999 \frac{\exp( - 4s)}{(4s)} - 2134 \frac{\exp( - 2.5s)}{(2.5s)} \nonumber \\
        -276 (1 - 0.005E / A_e ) \delta(s)
\label{seqn4}
\end{eqnarray} 
\noindent
where $E$ and $A_e$ are the laboratory energy and projectile mass number respectively. In the present case of radioactivity, it can be shown that $E/A_e=Qm/\mu$ where $m$ and $\mu$ are the nucleonic mass and reduced mass of the $A_e+A_d$ system, respectively, in units of MeV/c$^2$. The zero-range potential represents the single-nucleon exchange term and the density dependence $g(\rho_1, \rho_2)$ = $C (1 - \beta\rho_1^{2/3}) (1 - \beta\rho_2^{2/3})$ takes care of the higher order exchange effects and the Pauli blocking effects. This density dependent M3Y effective NN interaction supplemented by the zero-range potential is used to determine the nuclear matter equation of state. The equilibrium density of the nuclear matter is determined by minimizing the energy per nucleon. The density dependence parameters have been fixed by reproducing the saturation energy per nucleon and the saturation density of spin and isospin symmetric cold infinite nuclear matter. Although the density dependence parameters for single folding can be determined from the nuclear matter calculations and used successfully for proton radioactivity \cite{Ba05} and scattering \cite{Gu05,Gu06}, the transition to double folding is not straightforward. The parameter $\beta$ can be related to mean free path in nuclear medium, hence its value should remain same $\sim 1.6 fm^2$ as obtained from nuclear matter calculations \cite{bcs08} while the other constant $C$ which is basically an overall normalisation constant may change. The value of this overall normalisation constant has been kept equal to unity which has been found $\sim 1$ from optimum fit to a large number of alpha decay lifetimes. The density-dependence of the effective projectile-nucleon interaction is found to be fairly independent of the projectile, as long as the projectile-nucleus interaction is amenable to a single-folding prescription. This argument can be further stretched to mean that, in a double folding model, the density-dependent effects on the nucleon-nucleon interaction can be factorized into a `target term' times a `projectile term' \cite{Sr83}. 

\noindent
\label{section3}

    The cluster preformation probability $P_0$ for $\alpha$ is close to unity, {\it e.g.} 1, 0.6, 0.35 for even-even, odd-even and odd-odd emitters \cite{bmp93} respectively. The results of the present theoretical calculations for few $\alpha$ emitters are listed in Table I. Same set of $\alpha$ and cluster emitters of reference \cite{Bh08} are chosen. The zero point vibration energies used in the present calculations are the same as that described in reference \cite{Po86} immediately after eqn.(4) for the $\alpha$ cluster and by eqns.(5) for the heavier clusters. The shell effects for every cluster radioactivity are implicitly contained in the zero point vibration energy due to its proportionality with the $Q$ value, which is maximum when the daughter nucleus has a magic number of neutrons and protons. Values of the proportionality constants of $E_v$ with $Q$ is the largest for even-even parent and the smallest for the odd-odd one. Other conditions remaining same one may observe that with greater value of $E_v$, lifetime is shortened indicating higher emission rate. 

\begin{table}[htbp]
\caption{\label{table1} Half lives of $\alpha$ decay obtained in the present calculation. The asterisk symbol represents theoretical values assuming $\alpha$ preformation probability to be unity.} 
\begin{ruledtabular}
\begin{tabular}{cccccccc}

Parent & Cluster & $Q$ (MeV)&$logT_{ex}$ & $logT^*_{th}$ &$P_0$  &$logT_{th}$ & $l$   \\ 
\hline
$^{212}$Po&$^4$He&8.954&-6.52&-7.01&1.00&-7.01 & 0\\ 

$^{213}$Po&$^4$He&8.536&-5.44&-5.65&0.60&-5.43 & 0\\ 

$^{214}$Po&$^4$He&7.833&-3.78&-4.06&1.00&-4.06 & 0\\ 

$^{215}$At&$^4$He&8.178&-4.00&-4.47&0.35&-4.01 & 0\\

\end{tabular} 
\end{ruledtabular}
\end{table}
\noindent

    The theoretical half lives of cluster radioactivity for very heavy nuclei are calculated assuming cluster preformation factor to be unity. Hence the preformation factors $P_0$ can be calculated as the ratios of the calculated half lives to the experimentally observed half lives and listed in Table II. This represents the preformation factor and may be considered as the overlap of the actual ground state configuration and the configuration representing the cluster coupled to the ground state of the daughter. $P_0$ for normal $\alpha$ emitters is close to unity \cite{bmp93}. Superheavy emitters being loosely bound than highly bound $\alpha$, $P_0$ is expected to be high and present calculations with $P_0$=1 provide excellent description \cite{prc06,prc07,scb07} of $\alpha$ decay for recently discovered superheavy nuclei. For weakly bound heavy cluster decay it is expected to be orders of magnitude less than unity. 

\begin{table}[htbp]
\caption{\label{table2} Half lives assuming preformation factor to be unity and corresponding preformation factors of cluster decay obtained in the present calculation. The asterisk symbol represents theoretical values assuming preformation factor to be unity and using normalisation of 0.7 for the nuclear potentials.} 
\begin{ruledtabular}
\begin{tabular}{cccccccc}

 Parent & Cluster &$Q$ (MeV)& $logT_{ex}$ & $logT^*_{th}$ & $logT_{th}$ & $-logP_0$ & $l$   \\ 
\hline
$^{221}$Fr&$^{14}$C&31.317&14.52&13.46&11.71& 2.81 & 3\\ 

$^{221}$Ra&$^{14}$C&32.396&13.39&13.23&11.47& 1.92 & 3\\ 

$^{222}$Ra&$^{14}$C&33.050&11.00&10.38&8.71& 2.29 & 0 \\ 

$^{223}$Ra&$^{14}$C&31.829&15.20&14.34&12.56& 2.64 & 4\\ 

$^{224}$Ra&$^{14}$C&30.540&15.92&15.12&13.34& 2.58 & 0\\ 

$^{225}$Ac&$^{14}$C&30.477&17.34&17.25&15.39& 1.95 & 4\\ 

$^{226}$Ra&$^{14}$C&28.200&21.34&20.22&18.35& 2.99 & 0\\ 

$^{228}$Th&$^{20}$O&44.720&20.72&21.02&18.90& 1.82 & 0\\ 

$^{230}$U&$^{22}$Ne&61.400&19.57&20.81&18.23& 1.34 & 0 \\ 

$^{230}$Th&$^{24}$Ne&57.571&24.64&25.07&22.58& 2.06 & 0\\ 

$^{231}$Pa&$^{24}$Ne&60.417&23.38&22.74&20.32& 3.06 & 1\\ 

$^{232}$U&$^{24}$Ne&62.310&20.40&20.67&18.30& 2.10 & 0\\ 

$^{233}$U&$^{24}$Ne&60.486&24.82&24.31&21.82& 3.00 & 2\\ 

$^{234}$U&$^{24}$Ne&58.826&25.25&25.97&23.38& 1.87 & 0\\ 

$^{233}$U&$^{25}$Ne&60.776&24.82&24.35&21.92& 2.90 & 2\\ 

$^{234}$U&$^{26}$Ne&59.466&25.07&26.27&23.84& 1.23 & 0\\ 

$^{234}$U&$^{28}$Mg&74.110&25.74&25.66&22.93& 2.81 & 0\\ 

$^{236}$Pu&$^{28}$Mg&79.670&21.67&21.38&18.80& 2.87 & 0\\ 

$^{238}$Pu&$^{28}$Mg&75.912&25.70&26.25&23.49& 2.21 & 0\\ 

$^{238}$Pu&$^{30}$Mg&76.824&25.28&25.85&23.22& 2.06 & 0\\ 

$^{238}$Pu&$^{32}$Si&91.190&25.30&26.21&23.30& 2.00 & 0\\ 

$^{242}$Cm&$^{34}$Si&96.509&23.15&23.82&21.16& 1.99 & 0\\ 

\end{tabular} 
\end{ruledtabular}
\end{table} 
\noindent

    The basic problem of the present theoretical study is that there is no guarantee that the experimentally observed decays proceed from the ground state of the parent nucleus to that of the daughter nucleus which is assumed in the present calculations. It is a fundamental difficulty associated with the decay of the odd mass parent nuclei or the odd mass emitted clusters. These nuclei may be decaying predominantly to a low-lying excited state of the daughter nucleus. When the exotic cluster is removed from the parent nucleus, the state of the core left over may be quite different from that of the ground state of the daughter nucleus, but rather similar to that of one of its excited states. Hence, decays would go preferentially to this particular excited state, and be strongly suppressed to the ground (and other) states. Present model is not microscopic enough to predict, a priori, which excited state may be most appropriate for this role. However, since the effects due to nuclear structure do not appear explicitly, present calculations can also provide the theoretical half lives for transitions other than ground state to ground state if only the spin-parities and the corresponding experimental $Q$ values are precisely known.

    A systematic study of even-even parent and daughter combinations for which centrifugal barrier is zero, it was shown \cite{Zh04} that $-logP_0$ can be fitted either to $c_1 Z_d Z_e + c_2$ or to $c_3 A_e + c_4$ where $Z_d$, $Z_e$ are the charge numbers of the daughter and emitted nuclei. The folded nucleus-nucleus interaction potentials were obtained using M3Y effective interaction without any density dependence and the resulting nuclear potentials were renormalized (RM3Y) by multiplying them with a factor of 0.55 \cite{Zh04}. It is worthwhile to mention here that due to attractive character of the M3Y forces the saturation conditions for cold nuclear matter is not fulfilled \cite{Kh97}. The realistic description of nuclear matter properties can be obtained only with the density dependent M3Y effective interaction. Using DDM3Y effective interaction the saturation conditions can be fulfilled and in turn the constants of density dependence can be obtained from the saturation conditions of the cold symmetric nuclear matter. Hence, in the present case the nuclear potentials are not renormalised but the same density dependent effective interaction DDM3Y is used whose density dependence was obtained from nuclear matter calculations based on Hartree or mean field assumption \cite{bcs08} and was used quite successfully for proton scattering \cite{Gu05,Gu06}, proton radioactivity \cite{Ba05} and $\alpha$ radioactivity \cite{prc06,prc07,scb07}. Present calculations yield constants $c_1$ = -0.38 $\times$ 10$^{-3}$, $c_2$ = 2.56 from a very poor quality of fit and $c_3$ = 0.028, $c_4$ = 2.10 from somewhat better fit, which are quite different from reference \cite{Zh04}. These results imply absolutely no dependence on $Z_d Z_e$ and very weak dependence on $A_e$. The preformation probabilities found for the range of clusters considered here are $\approx$ 10$^{-2}$ to 10$^{-3}$ which tend to disagree with the results of a potential model with preformed clusters \cite{Bu89,Bu90,Bu94}. It is interesting to see in Table I that for $\alpha$ cluster the agreement is excellent. Although present calculations provide excllent estimates for the $\alpha$ decay half lives without adjusting the depth of the nuclear potentials, it is interesting to note that the theoretical cluster radioactivity half lives ($T^*_{th}$) are in reasonable agreement with experimental data spanning about fifteen orders of magnitude if a renormalisation of 0.7 for the nuclear potentials is used.

\noindent
\label{section4}

    To summarise, the half lives for cluster radioactivity have been calculated with microscopic nuclear potentials which are based on profound theoretical basis. Present calculation is found to be highly successful for $\alpha$ radioactivity including that of superheavies. For cluster radioactivity, present calculations with preformation probability $P_0=$ 1 systematically underestimate half lives, typically by factors of 10$^{-2}$ to 10$^{-3}$. The cluster preformation factors $P_0$ extracted from the ratio of calculated to measured half lives of cluster radoactivity shows very weak dependence on the mass number of the emitted nucleus and is roughly constant over the limited range of clusters (6$\le Z_e \le$14) considered here. The implication of such a state of affairs suggests contrary to all previous expectations, that it may be possible to extract nuclear structure information from exotic decay data. There seems to be a distinct chance that the single-particle orbitals filled by the daughter (core) nucleons are distorted and change their ordering due to polarisation effects caused by the presence of a large cluster. This, in turn, influences which excited state of the daughter nucleus is predominantly populated. It is interesting to note that the theoretical cluster radioactivity half lives ($T^*_{th}$) are in reasonable agreement with experimental data spanning about fifteen orders of magnitude if a renormalisation of 0.7 for the microscopic nuclear potentials is used. Many questions remain to be answered concerning the universality of this phenomenon, and whether different clusters cause varying amounts of core polarisation, so that different excited states of the same daughter nucleus are preferred when different clusters are emitted, and offers a very interesting challenge to more microscopic models, and to the experimentalists.

\end{document}